# SUPERHEATED DROPLET DETECTORS AS CDM DETECTORS: THE SIMPLE EXPERIMENT

J.I. COLLAR [§] [•], T.A. GIRARD [•], D. LIMAGNE [*], G. WAYSAND [*]

[§] *PPE Division, CERN, Geneve 23, CH-1211*

[•] *Centro de Física Nuclear, Universidade de Lisboa*
*1699 Lisboa, Portugal*

[*] *Groupe de Physique des Solides (URA 17 CNRS), Universités Paris VII/VI*
*75251 Paris, France*

Superheated Droplet Detectors (SDDs) are becoming commonplace in neutron personnel dosimetry. Their total insensitivity to minimum ionizing radiation (while responsive to nuclear recoils of energies ~ few keV), together with their low cost, ease of production, and operation at room temperature and 1 atm makes them ideal for Cold Dark Matter (CDM) searches. SDD's are optimal for the exploration of the spin-dependent neutralino coupling due to their high fluorine content. The status of SIMPLE (Superheated Instrument for Massive ParticLe Experiments) is presented. Under realistic background considerations, we expect an improvement in the present Cold Dark Matter sensitivity of 2-3 orders of magnitude after ~1 kg-y of data acquisition.

There is an evident need for large-mass, low-background detectors able to pervade the neutralino region of the WIMP phase space: the expected counting rate in most target materials is <1 / kg-day, and in order to detect the characteristic ~5 % annual modulation in the WIMP signal [1] within a reasonable time span, detector masses of order >50 kg are needed.

In this regard, Superheated Droplet Detectors (SDD) have been recently proposed as a convenient approach [2,3]. These devices consist of a dispersion of droplets of a superheated liquid (diameter ~10-100 μm), fixed in a viscous polymer or aqueous gel. SDDs are sensitive only to high-Linear Energy Transfer (LET) radiation and therefore energetic muons, gamma rays, x-rays and beta particles fall well below their activation threshold, typically >200 keV/μm. Strong irradiations show that SDDs are totally insensitive to gamma rays of energies <6 MeV at operating temperatures <30° C [4], while responding to neutron recoils of only a few keV. The dense energy deposition by a high-LET particle in a droplet breaks the metastability, causing its vaporization. This process is independent of the size of the droplet, a major advantage over Superheated Superconducting Granules (SSG). The resulting bubble (diameter ~1 mm) can be optically recorded and/or the characteristic (in frequency and time duration) sound emitted during the violent vaporization can be picked up [5]. SDD production involves the use of a high pressure reactor, but these detectors operate in a passive and safe fashion at room temperature and 1 atm, and have found great acceptation as neutron pocket dosimeters. While based on the same physical principles, the problems inherent to using a bubble chamber for dark matter detection [6] (pressure control, safety, duty-time and inhomogeneous nucleation) are absent from SDDs.

The following is a brief summary of SDDs' advantages for WIMP detection:

• Total insensitivity to low-LET radiation.

• Room temperature and atmospheric pressure continuous and safe operation.

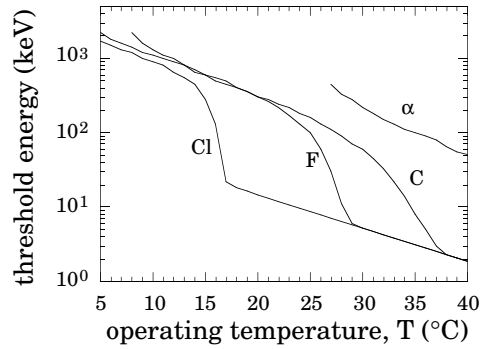

Figure 1: Bubble nucleation threshold energy for recoils from each of the atomic constituents in a freon-12 SDD, as a function of operating temperature. Alpha particles can induce nucleation only at T > 27°C [2,10].

• The low energy threshold for nuclear recoils can be tuned by temperature control (fig. 1). This is, for instance, ~5 keV at 30°C, ~16 keV at 20°C for freon-12 SDDs [7]. A differential spectrum is possible by temperature ramping.

• Known response to neutron recoils: monochromatic neutron calibrations have been performed [8,9] and are in good agreement with theoretical predictions (fig. 2). Similar predictions can be made for CDM candidates [2].

• Low cost of materials and shielding: less than 200 US $ per kg of active mass for "homemade" SDDs (~2,500 times more for commercial SDDs). Large target masses are a realistic possibility.

• Longevity and stability of response (measured for > 3 yr [11]).

• Simple optical and/or sound recording of bubble nucleation.

• SDDs are fluorine-rich (31% w.f. in freon-12), fluorine having the largest spin-dependent neutralino cross section [12].

• The freon-12 and -134a response falls dramatically for neutron energies <400 keV. This facilitates an efficient ambient neutron background reduction with modest water shielding. This water shielding conveniently doubles as a detector temperature regulator (water is circulated by a bath-temperature controller).

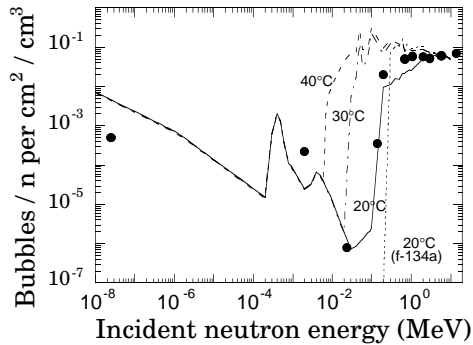

Figure 2: A comparison of our preliminary neutron response calculations for f-12 and f-134a with the experimental points at 20°C of [8]. Freon-134a has no response below 100 keV due to the absence of Cl in its composition [8].

• No heavy elements are required in the shielding, due to the SDD's inherent lack of response to photons. As a result, as far as the neutron background is concerned, only a shallow (~100 m.w.e.) underground site is required (fig. 3). Recoils directly induced by muons may require either active shielding or a slightly deeper site [13].

• Other applications such as solar and stellar collapse neutrino detection can be envisioned, taking advantage of the coherent nuclear elastic scattering cross section. We are exploring the possible use of other refrigerant liquids for this purpose (fig. 4).

Even for such an advantageous type of detector, some backgrounds must be considered and reduced [2]:

§ SDDs are highly stable against mechanical vibrations [16].

§ Homogeneous (spontaneous) nucleation is negligible for the envisioned degrees of superheat. According to the classical theory of homogeneous nucleation [17], a temperature of > 91°C (at 1 atm) would be needed to produce a background of concern in a CDM search.

§ Neutrons from natural radioactivity in rock and detector materials: we have performed MCNP-4a Monte Carlo simulations using the Gran Sasso ambient neutron spectrum as an input, showing that ~40 cm of hydrogenated shielding are enough to reduce this background to ~0.01 bubbles/kg freon-12/day. The required radiopurity in the shielding itself is <0.01 ppb U/Th, a level found even in some US tap water.

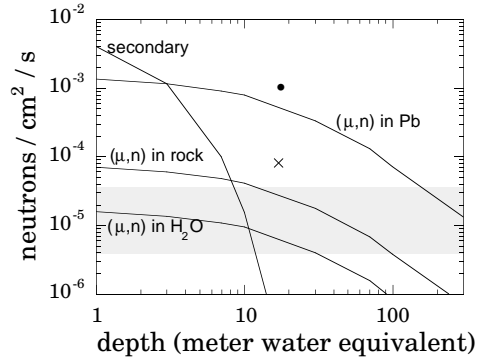

Figure 3: Components of neutron flux as a function of underground depth. The greyed band covers the typical contribution from natural radioactivity in rocks, based on large depth measurements and the average U and Th concentration in the crust. The contribution to the flux from muon interactions is derived from the systematics and measurements of [14]. In the absence of high-Z shielding, there is no significant reduction in flux below ~100 m.w.e.. Many urban locations are available at such shallow depths. The • (x) marks the Pb-shielded (unshielded) flux in the Stanford underground facility. Similar calculations are found in [15].

§ Neutrons from (μ,n) in shielding and detector are unimportant below few tens of m.w.e. (fig. 3).

§ While α particles cannot produce nucleation below ~27°C (fig. 1), the accompanying α-recoils can. A radiopurity in the superheated liquid of <$10^{-6}$ ppb is needed to render this background less important than those listed above. Fractional distillation of the freon, if needed, is a possibility. For low filling factors, the short range of the α-recoils (~0.1 μm) makes the nucleation efficiency for gel-originated α-recoils very small; <0.01 ppb U/Th in the gel are sufficient.

§ Auger electrons are high-LET particles, yet the maximum energy deposition for freon-12 is ~2.5 keV, corresponding to the binding energy of the K-shell in Cl and below threshold for this search (fig. 1).

Taking the moderated neutron background as dominant, it is shown [2] that an improvement of 2-3 orders of magnitude in the present spin-independent WIMP sensitivity is expected from a 1 kg active mass SDD after only 1 year of data acquisition, widely exploring the expected neutralino phase-space.

It is evident that optimization of commercial SDDs must be made, for instance, in order to meet the background requirements listed above. For this reason, and foreseeing the construction of inexpensive multi-kg SDDs, a large (97 cm length, 30 cm diam.) pressure reactor, dedicated to SDD production, was built in cooperation with COMEX-PRO. The reactor withstands 60 bar (allowing us to experiment with a large variety of superheated liquids) and features two large inspection windows and a 16 l tray that can be used to collect the SDD. The reactor presently houses two Ismatec MS-A31016 biological pumps, with external flow controls, used to propel freon and gel during production. We have used a coaxial-flow method of production [18] where freon is injected through a small (200 µm) nozzle formed by electro-erosion. This freon jet is surrounded by a gel flow, the laminar flow is broken spontaneously, and small freon droplets are formed. We have recently installed an electromagnetic stirrer inside the chamber, which allows for the simpler and widely used method of making a freon and gel "mayonnaise". So far we have only used gelatine, glycerol and Aquasonic [19] -based gels, but we intend to experiment with polyvinyl alcohol (PVA), polybutylene and acrylamide preparations that should yield a more stable product.

A shallow underground location (26 m calcite overburden), with convenient horizontal access has been found 40 km south of Paris. Measurements of the radiopurity of the rock and tarmac have been performed and we will soon start measurement of the (paraffin shielded) ambient neutron flux using commercial SDDs. In a first prototype-stage experiment, to start immediately, we plan to install a ~100g active mass SDD in a 30 lt water bath, held stable within 0.1°C by a Bioblock Polystat temperature-controller. The bath will be surrounded by ~35 cm paraffin, already available. The sound emission produced by the bubble vaporization will be picked-up by electret microphones, amplified, and the pulse shape registered in a LeCroy 9350 AM digital storage oscilloscope. Our first objectives are to determine the intrinsic radiopurity of the freon liquid, to measure the dominant backgrounds and to characterize the detector using a weak Cf-252 neutron source (we also intend to instrument a few commercial SDDs in the same shielding for comparison purposes). It is however interesting to note that in the absence of unforeseen backgrounds, limits comparable to those from ultra-low background Ge detectors could be obtained in ~2 mo. even with such a small-mass prototype. The next stage will involve the installation of a multi-kg SDD in the same or a deeper location, depending on the prototype results.

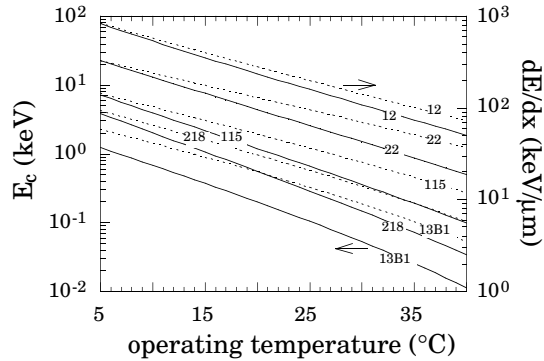

Figure 4: Minimum energy deposition and stopping power required for bubble nucleation in some refrigerant liquids (freon-x, labelled by x). Some other suitable moderately superheated refrigerants are R-134a, -143a, -32, -125. Possible target liquids contain heavy elements such as I and Br and may be envisioned as solar neutrino detectors. As an indicative value, a B-8 solar neutrino can impart a maximum kinetic energy of ~14 keV to a Cl nucleus.

**References:**

1. A.K. Drukier *et al.*, Phys. Rev. D **33**, 3495 (1986) .
2. J.I. Collar, Phys. Rev. D **54**, R1247 (1996).
3. L.A. Hamel *et al.*, Proc. of 2nd Workshop on The Dark Side of the Universe, Rome, Italy, November 1995, prep. hep-ex/9602004.
4. W. Lim *et al.*, Nucl. Instr. Meth. A **370**, 568 (1996);
S.G. Vaijapurkar *et al.*, Radiat. Meas. **24**, 309 (1995) and Radiat. Meas. **23**, 753 (1994); M.J. Harper *et al.*, Health Physics **68**, 670 (1995); C.A. Perks *et al.*, Radiat. Prot. Dosim. **23**, 131 (1988); H. Ing *et al.*, Nucl. Tracks Radiat. Meas. **8**, 285 (1984).
5. R.E. Apfel *et al.*, Rev. Sci. Instrum. **54**, 1397 (1983).
6. V. Zacek, Nuovo Cimento A **107**, 291 (1994).
7. Freon-12 is a registered trademark of the Du Pont Co.
8. Y.-C. Lo *et al.*, Phys. Rev. A **38**, 5260 (1988).
9. M.J. Harper, Ph. D. dissertation, University of Maryland, 1991
10. C.K. Wang *et al.*, Nucl. Instr. & Meth. A **353**, 524 (1994).
11. J.A. Sawicki, Nucl. Instr. & Meth. A **336**, 215 (1993).
12. J. Ellis, Nuovo Cimento A **107**, 1091 (1994).
13. P.B. Price, private communication.
14. G.V. Gorshkov *et al.*, Sov. J. Nucl. Phys. **13**, 450 (1971).
15. G. Heusser, Nucl. Instr. Meth. A **369**, 539 (1996).
16. T. Cousins *et al.*, IEEE Trans. Nucl. Sci. **37**, 1769 (1990).
17. M. Blander *et al.*, Am. Inst. Chem. Eng. J. **21**, 833 (1975).
18. Y.-C. Lo, Ph. D. dissertation, Yale University, 1987.
19. Aquasonic is a registered trademark of Parker laboratories, NJ.